\documentclass[aps,prb,reprint,twocolumn,floatfix,superscriptaddress,longbibliography]{revtex4-2}

\usepackage{standalone}
\usepackage{amsthm}
\usepackage{lipsum}
\usepackage{amsfonts, amsmath, amssymb}
\usepackage{graphicx}
\usepackage{dcolumn}
\usepackage{bm}
\usepackage{dsfont}
\usepackage[normalem]{ulem}
\usepackage{color}
\usepackage[utf8]{inputenc}
\usepackage[colorlinks,citecolor=blue,linkcolor=blue,urlcolor=blue]{hyperref}
\usepackage{tikz}
\usepackage{braket}
\usepackage{physics}
\usepackage{color}
\usepackage{soul}
\usepackage{mathtools}
\usepackage{float}
\usepackage{esvect}
\usepackage{subfigure}
\usepackage{stmaryrd} 
\usepackage{enumerate}
\usepackage{bbold}

\newcommand{\SM}{Appendix}

\makeatletter
\def\maketitle{
	\@author@finish
	\title@column\titleblock@produce
	\suppressfloats[t]}
\makeatother

\begin{document}

\title{Strange Luttinger liquids in a cavity-embedded one-dimensional electronic chain}
\author{Danh-Phuong Nguyen}
\affiliation{Department of Physics, University of Konstanz, 78464 Konstanz, Germany}
\affiliation{Universit\'{e} Paris Cit\'e, CNRS, Mat\'{e}riaux et Ph\'{e}nom\`{e}nes Quantiques, 75013 Paris, France}
\author{Christophe Mora}
\affiliation{Universit\'{e} Paris Cit\'e, CNRS, Mat\'{e}riaux et Ph\'{e}nom\`{e}nes Quantiques, 75013 Paris, France}
\author{ Cristiano Ciuti}
\affiliation{Universit\'{e} Paris Cit\'e, CNRS, Mat\'{e}riaux et Ph\'{e}nom\`{e}nes Quantiques, 75013 Paris, France}

\begin{abstract}
\noindent 
We study a one-dimensional electronic chain coupled to a homogeneous quantized vacuum field and electron–electron interactions. In the absence of the latter, we derive a low-energy effective description in the presence of light–matter coupling, which we identify as a \textit{strange Luttinger liquid}. Although it retains a formal resemblance to conventional Luttinger liquid theory, the coupling to the quantum field qualitatively modifies the low-energy sector and breaks the standard velocity relation underlying Luttinger universality. For finite electron-electron interactions, we recover a phase diagram featuring several phases as a function of interaction strength and hopping amplitude, including a phase hosting Majorana-like zero modes. Using exact diagonalization, we compute observables that characterize the phase boundaries and show that the cavity field significantly shifts them. We also study the fate of Majorana-like states under the influence of the cavity field, highlighting their modification by light–matter coupling. Finally, we investigate whether the strange Luttinger liquid description identified in the noninteracting regime continues to hold when electron–electron interactions are introduced.

\end{abstract}
\date{\today}
\maketitle
\section{Introduction}
In recent years, strong cavity vacuum fields have emerged as a powerful means of controlling the properties of condensed-matter systems \cite{GarciaVidal2021,Schlawin2022,Bloch2022,Lu2025,Basov2025}. Experimental platforms based on metallic split-ring terahertz resonators \cite{Scalari2012,Keller2017,ParaviciniBagliani2018,Enkner2025} and hyperbolic van der Waals materials \cite{Ashida2023} have enabled access to this regime. These implementations have revealed signatures of strong light–matter coupling in magnetotransport measurements in the Shubnikov–de Haas regime \cite{ParaviciniBagliani2018}, modifications of the integer quantum Hall effect in both two-dimensional electron gases \cite{Appugliese2022,Kuroyama2023,Kuroyama2024} and graphene \cite{Xue2025}, enhancements of fractional quantum Hall gaps \cite{Enkner2025}, and shifts of the critical temperature associated with charge-density-wave order \cite{Jarc2023}. Collectively, these results demonstrate the capacity of cavity fields to engineer correlated electronic phases. On the theoretical side, a variety of models have been developed to investigate the influence of vacuum electromagnetic fields on superconductivity \cite{Sentef2018, Curtis2019, Schlawin2019, Kozin2024, Gomez-Leon2024}, quantum transport \cite{Hagenmller2017, Hagenmller2018, Arwas2023, Rokaj2023, Winter2025, Macedo2024, Borici2025}, and topological phenomena \cite{Dmytruk2022, Li2022, Shaffer2023,Lin2023,Dmytruk2024,Dag2023,PerezGonzalez2025,MndezCrdoba2020,MndezCrdoba2023,Yang2024,Becerra2025, Buonemani2026, Ritz-Zwilling2026, Santos2026, Shin2026, Yang2026, Bacciconi2025, Xavier2025}. Notably, the topology of hybridized light–matter states with finite photonic weight is beginning to attract substantial attention, suggesting the emergence of new topological classifications tailored to cavity-embedded systems \cite{Nguyen2023,Nguyen2024,Bacciconi2024,Karle2025}.

Over the past decade, the pursuit of quantum computing has accelerated across multiple experimental platforms \cite{Antonio2018,Proctor2025}. Among the proposed implementations, Majorana bound states in topological superconductors \cite{Kitaev2001, Alicea2012, Beenakker2013, Prada2020} remain particularly promising \cite{Sarma2015}, owing to their robustness against \textit{local} sources of disorder. More recently, attention has turned to how these states respond to \textit{global} perturbations such as cavity fields, which have been the focus of extensive investigation \cite{Dmytruk2024, Gomez-Leon2024, Bacciconi2024, Becerra2025}. Nevertheless, engineering strong light–matter coupling in systems hosting Majorana modes, or in superconductors in general, presents significant challenges. Standard approaches often rely on breaking particle-number conservation, or equivalently the global $U(1)$ symmetry. To introduce coupling to quantized vacuum fields, one typically employs gauge transformation in both proximity-induced and intrinsic superconductors \cite{Dmytruk2024, Gomez-Leon2024, Becerra2025}, or derives the light–matter interaction from a fully microscopic description in which the pairing terms of an intrinsic superconductor are determined self-consistently \cite{Schirmer2022, Kozin2025}. However, non-conservation of particle numbers is not the only framework within which superconductivity can be understood. Leggett’s number-conserving variational ansatz reproduces the conventional gap equation while preserving total particle number \cite{Leggett2006}, and this perspective has motivated skepticism regarding the realization of Majorana modes in superconductors intended for quantum computation \cite{Lin2022}. Motivated by this viewpoint, several works have constructed number-conserving models that host topological superconducting phases \cite{Kraus2013, Chen2018, Iemini2016, Vadimov2021}. In this spirit, we investigate light–matter interactions in a one-dimensional chain with nearest-neighbor electron–electron interactions, representing a mechanism for intrinsic superconductivity, although the superconducting aspects of the model will not be the primary focus of this work.

The paper is organized as follows. In Sec. \ref{sec: micro H}, we introduce the microscopic Hamiltonian describing a tight-binding chain with nearest-neighbor electron–electron interactions and light–matter coupling. While each of these ingredients has been studied separately in previous works \cite{Vadimov2021,Eckhardt2022}, their combined effect within a unified model has not been addressed to the best of our knowledge. We then briefly review the Luttinger liquid framework in the limit of vanishing light–matter coupling. In Sec. \ref{sec: CLL}, we focus on the case without electron–electron interactions and derive an effective low-energy Hamiltonian for the system. Although this derived model shares similar structure with conventional Luttinger liquids, we show that it does not obey their standard relation between velocities, and instead realizes a \textit{strange} Luttinger liquid regime. Finally, to capture both two-body interaction and cavity effects, Sec. \ref{sec: ED} presents first exact diagonalization results for finite systems. We analyze the local density of states and a many-body topological marker as signatures of Majorana-like zero modes \cite{Vadimov2021}, as well as charge susceptibilities characterizing the charge-density-wave and phase-separation transitions \cite{Haldane1980,Giamarchi2003,Sutherland2004}, all in the presence of light–matter coupling. We further examine whether the strange Luttinger liquid framework remains valid in the interacting regime. Conclusion and Outlook will be discussed in Sec. \ref{sec: conclusion}.

\section{Microscopic Hamiltonian \& conventional Luttinger Liquids}
\label{sec: micro H}
Throughout this work, we set $\hbar = a = e =1$, where $\hbar$ is the reduced Planck constant, $a$ is the lattice spacing, and $e$ is the elementary charge. We study a cavity-embedded tight-binding chain consisting of $N$ sites under periodic boundary conditions. The Hamiltonian of this system can be written as:
\begin{equation}
\label{eq: micro H}
    \begin{aligned}
        \hat{H} &= \omega_c\hat{a}^{\dagger}\hat{a}-\left[te^{-ig(\hat{a}+\hat{a}^{\dagger})}\sum_{n=1}^{N}\hat{c}^{\dagger}_{n+1}\hat{c}_n + \text{h.c.}\right] \\
        &+ U\sum_{n=1}^{N}\left(\hat{c}^{\dagger}_{n+1}\hat{c}_{n+1}-\frac{1}{2}\right)\left(\hat{c}^{\dagger}_{n}\hat{c}_{n}-\frac{1}{2}\right).
    \end{aligned}
\end{equation}
In the above expression, the matter part is described as follows: $\hat{c}_n$ ($\hat{c}^{\dagger}_n$) annihilates (creates) an electron at the site $n$, $t > 0$ denotes the hopping coefficient between neighboring sites, and $U$ represents the electron-electron interaction strength. The $-1/2$ term is introduced to preserve particle-hole symmetry. For the light part, we consider a homogeneous, single-mode cavity, where $\hat{a}$ ($\hat{a}^{\dagger}$) denotes the annihilation (creation) operator of a single photon with frequency $\omega_c$. The light-matter interaction is incorporated via Peierls substitution, resulting in $t \rightarrow t e^{-ig(\hat{a}+\hat{a}^{\dagger})}$, where $g$ is the dimensionless coupling strength. This study focuses on a large but finite chain whose length remains much smaller than the cavity mode volume. The corresponding results in the thermodynamic limit can be obtained through a simple rescaling $g \rightarrow g/\sqrt{N}$. 

Let us first consider the limiting case: $g = 0$. In the absence of light-matter coupling, Eq. (\ref{eq: micro H}) has been extensively studied in the literature \cite{Giamarchi2003, Sutherland2004}. In particular, its low-energy many-body excitation spectrum can be described within the framework of LL theory \cite{Haldane1981, Bouchoule2025}. Taking a non-interacting Fermi sea $\vert \psi_0\rangle$ with $N_0$ electrons (assumed even) as a reference state, this approach amounts to restricting the physics to a narrow momentum window $\Lambda$ around the Fermi points, namely $\vert k - \sigma k_F\vert < \Lambda$, where $\sigma = \pm$ labels right- and left-moving fermions, and $k_F = \pi N_0/N$, one type of excitation corresponds to adding extra particles near the two Fermi points. Denoting by $N_{\sigma}$ the number of electrons added around $\sigma k_F$, the total number of particles becomes $N_e = N_0 + N_+ + N_-$. The imbalance between the two branches is given by $J = N_+ - N_-$, which is proportional to the persistent current. $N_e$ and $J$ satisfy the parity constraint $(-1)^{N_e} = - (-1)^{J}$, and they remain good quantum numbers of $\hat{H}$ \cite{Haldane1981,Giamarchi2003,Sutherland2004}. The second type of excitation consists of particle-hole pairs generated in the vicinity of the Fermi points. Owing to the linearity of the single-particle spectrum in this regime, these excitations are degenerate in energy and can therefore described by a collective excitation $\hat{\rho}_{q,\sigma} = \sum_{\vert k - \sigma k_F\vert < \Lambda}\hat{c}^{\dagger}_{k+q}\hat{c}_{k}$ for $q = n 2\pi/N$ with non-zero integer $n$. In the non-interacting limit $U = 0$, the bosonic modes are introduced through $\hat{b}^{\dagger}_q = \sqrt{2\pi/(N\vert q\vert)} \sum_{\sigma = \pm}\theta(\sigma q)\hat{\rho}_{q,\sigma}$, where $\theta(x)$ is the Heaviside step function. Defining $\Delta N_e = N_e - N_0$, $\epsilon_F = -2t\cos k_F$, $v_F = +2t\sin k_F$, the low-energy spectrum of $\hat{H}$ is described by the Luttinger Hamiltonian $\hat{H}_{LL}$ \cite{Haldane1981}:
\begin{equation}
    \label{eq: TLL}
    \begin{aligned}
        \hat{H}_{LL}(v_S,v_N,v_J) &= \epsilon_F \Delta N_e + v_S\sum_q \vert q\vert \hat{b}^{\dagger}_q\hat{b}_q\\
        &+ \frac{\pi v_N}{2N}\Delta N_e^2 + \frac{\pi v_J}{2N}J^2.
    \end{aligned}
\end{equation}
Here, $v_N$ and $v_J$ represent the propagation velocities of density and current fluctuations, while $v_S$ defines the actual physical speed of sound for collective excitations. For a non-interacting chain, we have $v_S = v_N = v_J = v_F$. For finite $U$, Eq. (\ref{eq: TLL}) continues to describe the low-energy physics. Although $\Delta N_e$ and $J$ remain good quantum numbers, the operators $\hat{b}^{\dagger}_q\hat{b}_q$ no longer are. Instead, the second type of elementary excitation are expressed by dressed bosonic operators $\hat{\beta}^{\dagger}_q$ and $\hat{\beta}_q$, related to $\hat{b}^{\dagger}_q$ and $\hat{b}_q$ through Bogoliubov transformation. Such electron-electron interaction also modify the three velocities, nevertheless they still satisfy the exact relation:
\begin{equation}
    \label{eq: equality}
    v_S = \sqrt{v_J v_N}.   
\end{equation}
This identity implies that low-energy properties of Luttinger liquid are governed by a \textit{single} dimensionless parameter $K = \sqrt{v_J/v_N}$, which not only determines the asymptotic behavior of the correlation functions but also the phase transition to charge-density-wave phase and phase separation \cite{Giamarchi2003, Sutherland2004}. Later in Sec. \ref{sec: CLL}, we will show that these relations are violated in the presence of the cavity. 

Additionally, when open boundary conditions are imposed on Eq. (\ref{eq: micro H}), Ref. \cite{Vadimov2021} reported the emergence of many-body Majorana-like zero modes at half-filling without invoking mean-field treatment, thereby conserving the $U(1)$ symmetry. To be specific, they employed both Density Matrix Renormalization Group (DMRG) calculations and bosonization techniques to demonstrate the presence localized edge states in an open chain. The term ``Majorana-like'' refers to the fact that these zero-energy modes exhibit a power-law scaling of the energy splitting with system size, consistent with a gapless bulk, in contrast to conventional Majorana modes that are protected by a finite energy gap. To recover the gapped bulk without resorting to mean-field treatment, previous works have considered double wire or ladder systems \cite{Kraus2013, Chen2018}. Nevertheless, the single-chain configuration offers a fundamental advantage for studying light-matter coupling exactly in the limit of vanishing electron-electron interaction, and perturbatively for finite $U$, as we will demonstrate later.

\section{Strange Luttinger Liquid Formalism}
\label{sec: CLL}
The central result of this Section is that, in the absence of the two-body interacting term ($U = 0$), each photon sector $n$ exhibits a many-body energy spectrum that can be described by a Hamiltonian $\hat{H}_n$ such that:
\begin{equation}
\label{eq: HnLL}
    \hat{H}_n \simeq \hat{H}_{LL}(\bar{v}^{(n)}_S,\bar{v}^{(n)}_N,\bar{v}^{(n)}_J),
\end{equation}
where $\hat{H}_{LL}$ retains the exact form of Eq. (\ref{eq: TLL}), while the effective velocities $\bar{v}_S^{(n)}$, $\bar{v}_N^{(n)}$ and $\bar{v}_J^{(n)}$ are renormalized by the cavity. Importantly, despite this formal similarity to the conventional conventional Luttinger liquid, we find that $\bar{v}_S^{(n)} \neq \sqrt{\bar{v}^{(n)}_J \bar{v}^{(n)}_N}$. As a consequence, Eq. (\ref{eq: equality}) no longer holds in each photon sector, signaling a breakdown of the conventional structure and motivating the notion of strange Luttinger liquid. In the following, we clarify the origin and implications of this violation. 

\subsection{Low-energy description of factorized states}
In the limit $U = 0$, the Hamiltonian (\ref{eq: micro H}) has been studied in Ref. \cite{Eckhardt2022}. Due to the homogeneity of the cavity mode, the Hamiltonian preserves discrete translational symmetry, allowing one to perform a Fourier transformation and rewrite Eq. (\ref{eq: micro H}) in the following form:
\begin{equation}
    \label{eq: Hk}
    \begin{aligned}
    \hat{H}_0 = \omega_c \hat{a}^{\dagger}\hat{a} + \text{cos}\left[g(\hat{a}+\hat{a}^{\dagger})\right]\hat{T} + \text{sin}\left[g(\hat{a} +\hat{a}^{\dagger})\right]\hat{I},
    \end{aligned}
\end{equation}
where the kinetic and current terms are given, respectively, by:
\begin{equation}
    \label{eq: T and I}
    \begin{aligned}
        \hat{T} &= -2t\sum_{k}\text{cos}k\:\hat{c}^{\dagger}_k\hat{c}_k \equiv \sum_k \epsilon_k \hat{c}^{\dagger}_k\hat{c}_k,\\
        \hat{I} &= +2t\sum_k \text{sin}k\:\hat{c}^{\dagger}_k\hat{c}_k \equiv \sum_k v_k \hat{c}^{\dagger}_k\hat{c}_k.
    \end{aligned}
\end{equation}
Eqs. (\ref{eq: Hk}) and (\ref{eq: T and I}) clearly indicate that a generic electronic occupation $\{n_k\}$ where $n_k = 0$ or $1$ denotes whether the state $k$ is occupied, remains a good quantum number for Eq. (\ref{eq: Hk}). Consequently, eigenstates of $\hat{H}_0$ can be expressed as $\vert \psi\rangle\vert \phi_{\psi}\rangle = \vert \{n_k\}\rangle\vert\phi_{\psi}\rangle$, implying the absence of light-matter entanglement. The photon state $\vert \phi_{\psi}\rangle$ is obtained by diagonalizing  the projected Hamiltonian $\langle \psi \vert \hat{H}_0\vert \psi\rangle$, thus the many-body light-matter problem is reduced to a many-body photon problem, which possesses a significantly smaller Hilbert space. As a result, the eigenstates and energies of Eq. (\ref{eq: Hk}) can be solved \textit{exactly} even for large systems. The ground state of this system remains the non-interacting Fermi sea $\vert \psi_0\rangle$, albeit with dressed energy \cite{Eckhardt2022}. 

After projecting Eq. (\ref{eq: Hk}) onto a generic occupied electronic state $\vert\psi\rangle = \vert\{n_k\}\rangle$, the corresponding photon state $\vert \phi_{\psi,n}\rangle$ must satisfy $\hat{H}_0^{(p)}(\omega_c,g,E,I)\vert \phi_{\psi,n}\rangle = \mathcal{E}_n(\omega_c,g,E,I)\vert\phi_{\psi,n}\rangle$, where $E = \langle \psi \vert \hat{T}\vert \psi \rangle$, $I = \langle \psi \vert \hat{I}\vert \psi \rangle$, and:
\begin{equation}
\label{eq: Hp}
    \hat{H}_0^{(p)} = \omega_c \hat{a}^{\dagger}\hat{a} + \text{cos}\left[g(\hat{a}+\hat{a}^{\dagger})\right]E + \text{sin}\left[g(\hat{a} +\hat{a}^{\dagger})\right]I.
\end{equation}
Note that the quantum number $n$ refers to the photon degree of freedom and can be interpreted as a \textit{dressed} number state for each specific pair of $E$ and $I$. In this work, we focus exclusively on low-energy excitations above the non-interacting Fermi sea $\vert \psi_0\rangle$. By defining $E_0 = \langle \psi_0 \vert \hat{T} \vert \psi_0\rangle$ and $I_0 = \langle \psi_0 \vert \hat{I}\vert \psi_0 \rangle = 0$, we expand $\mathcal{E}_n$ around $E_0$ and $I_0$, yielding $\mathcal{E}_n(\omega_c,g,E,I) \simeq \mathcal{E}_n(\omega_c,g,E_0,0) + \lambda_1^{(n)}(E -E_0) + \lambda_2^{(n)} (E-E_0)^2/2 + \eta_2^{(n)}I^2/2$, where the coefficients are defined as follows:
\begin{equation}
\label{eq: 4 coeffs}
    \begin{aligned}
        &\lambda_1^{(n)} = \frac{\partial \mathcal{E}_n}{\partial E}(\omega_c,g,E_0,0), \: \lambda_2^{(n)}= \frac{\partial^2\mathcal{E}_n}{\partial E^2}(\omega_c,g,E_0,0),\\
        &\eta_1^{(n)} = \frac{\partial \mathcal{E}_n}{\partial I}(\omega_c,g,E_0,0), \: \eta_2^{(n)}= \frac{\partial^2\mathcal{E}_n}{\partial I^2}(\omega_c,g,E_0,0).
    \end{aligned}
\end{equation}
Eq. (\ref{eq: 4 coeffs}) can be calculated using the Hellmann - Feynman theorem, which depends only on the eigenstates of the Hamiltonian (\ref{eq: Hp}) at $(\omega_c,g,E_0,0)$. Moreover, without performing the explicit calculation, we note that $\mathcal{E}_n(I)$ is symmetric around 0, since changing the sign of $I$ is equivalent to a gauge transformation $\hat{a} \rightarrow -\hat{a}$, which leaves the physical properties unchanged. For this reason, $\eta_1^{(n)} = 0$, and the energy expansion in $I$ must start at second order. Since a generic occupied electronic state $\vert \psi \rangle = \vert \{n_k\}\rangle$ is an eigenstate of $\hat{H}_0$, $\hat{T}$ and $\hat{I}$, we can replace $E$, $I$ by the operators $\hat{T}$, $\hat{I}$ in the energy expansion. This yields an effective electronic Hamiltonian corresponding to the dressed $n$-photon subspace:
\begin{equation}
    \label{eq: micro Hn}
    \begin{aligned}
        \hat{H}_n &= \mathcal{E}_n(\omega_c,g,E_0,0) + \lambda_1^{(n)}(\hat{T}-E_0)\\
        &+ \frac{\lambda_2^{(n)}}{2}(\hat{T}-E_0)^2 + \frac{\eta_2^{(n)}}{2} \hat{I}^2.
    \end{aligned}
\end{equation}
Before proceeding further, we first discuss the scaling of the coupling constants introduced in Eq. (\ref{eq: 4 coeffs}). In systems with many electrons such that $\vert E_0 \vert \gg \omega_c$, the sine and cosine functions in Eq. (\ref{eq: Hp}) can be expanded to leading order \cite{Kozin2025}. In this limit, $\lambda_1^{(n)}$, $\lambda_2^{(n)}$ and $\eta_2^{(n)}$ can be analytically determined via a Bogoliubov transformation, with explicit formulas provided in the \SM. Here we only summarize the main results. First, $\lambda_1^{(n)} \leq 1$, and $\lambda_2^{(n)}$, $\eta_2^{(n)} \leq 0$, with equality holding when $g=0$. The $\lambda_1^{(n)}$ governs the squeezing of the band or mass renormalization \cite{Rokaj2022}, while $\eta_2^{(n)}$ ($\lambda_2^{(n)}$) is related to the current-current (charge-charge) correlator \cite{Eckhardt2022}. More importantly, we observe that:
\begin{equation}
\label{eq: scaling}
    \frac{\lambda^{(n)}_2}{\lambda_1^{(n)}} = \mathcal{O}(N^{-3/2}),\: 
    \frac{\eta^{(n)}_2}{\lambda_1^{(n)}} = \mathcal{O}(N^{-1}). 
\end{equation}
which is useful for simplifying the effective Hamiltonian by retaining only the dominant terms.

\subsection{Derivation of strange Luttinger liquid}
In the next step, we aim to describe the low-energy physics of Eq. (\ref{eq: micro Hn}). The bosonized representations of Eq. (\ref{eq: T and I}) read:
\begin{equation}
    \begin{aligned}
    \hat{T} &\simeq \epsilon_F \Delta N_e + v_F\sum_q \vert q\vert \hat{b}^{\dagger}_q\hat{b}_q
        + \frac{\pi v_F}{2N}\left(\Delta N_e^2 + J^2\right),\\ 
    \hat{I} &\simeq N\frac{\partial \hat{T}}{\partial \phi}\Big{|}_{\phi=0} = -v_F J.
    \end{aligned}
\end{equation}
Here, the kinetic term $\hat{T}$ takes the same form as in Eq. (\ref{eq: TLL}) where all velocities equals $v_F$, while the current counterpart can be obtained by threading a flux $\phi$ through the system, leading to $J \rightarrow J - \phi/\pi$.
\begin{figure}[t!]
\includegraphics[width = 1.0\hsize]{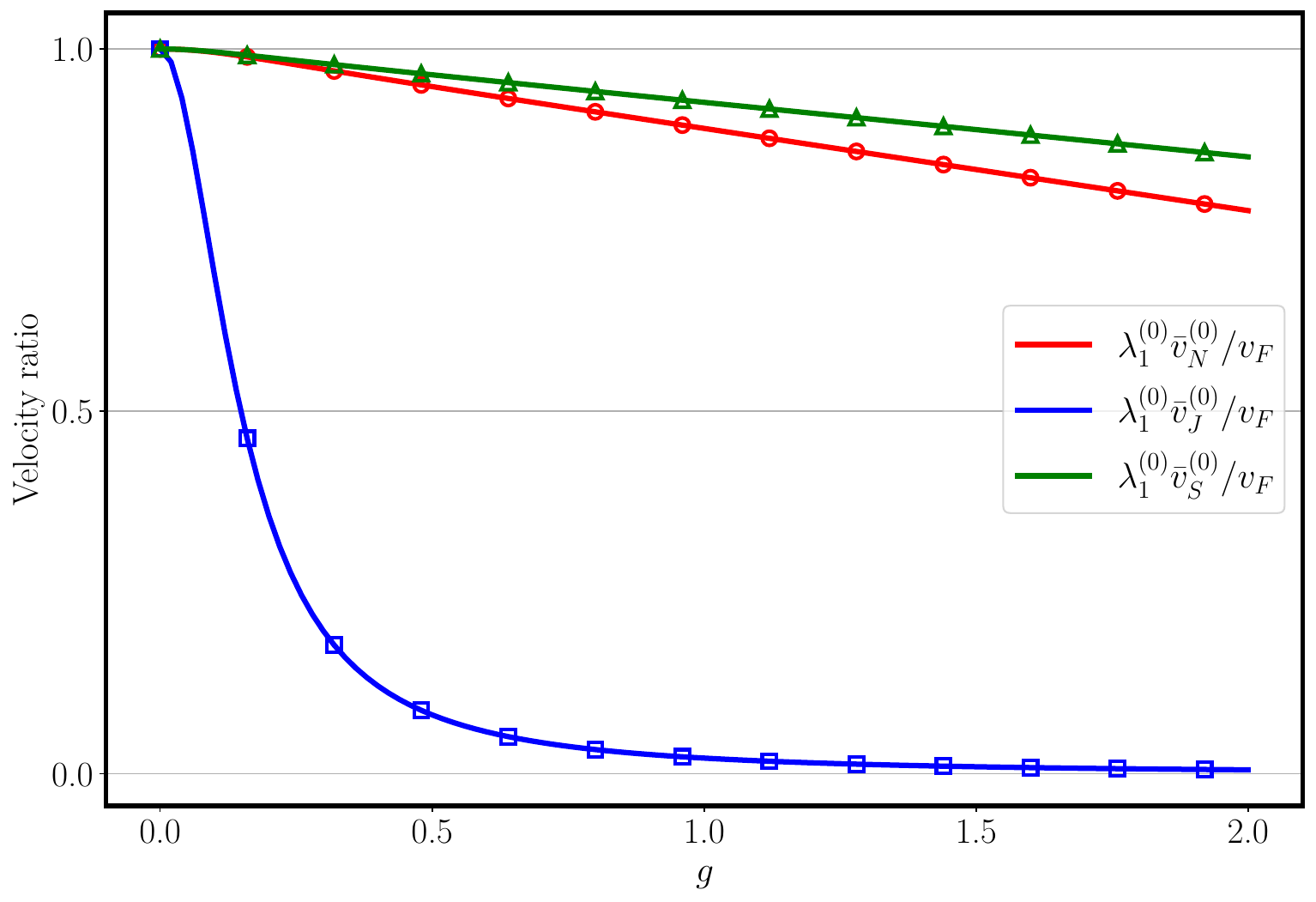}
  \caption{Characteristic velocities of the strange Luttinger liquid, normalized by the Fermi velocity, as functions of the cavity coupling $g$. Solid lines indicate values obtained from Eq. (\ref{eq: 3 velocities g}), while open symbols correspond to exact velocities calculated directly from the tight-binding model given by Eq. (\ref{eq: Hk}).  The sound velocity of the collective excitations is given by $\bar{v}_S^{(0)}$, while $\bar{v}_N^{(0)}$ and $\bar{v}_J^{(0)}$ characterize the propagation velocities of density and current fluctuations, respectively. The parameter $\lambda_1^{(0)}$ quantifies the cavity-induced band squeezing. Other parameters: $N = 500$, $N_e/N = 1/4$, $\omega_c/t = 10$.
}
\label{manu_fig: 3v}
\end{figure}
Detailed calculations are provided in the \SM. Consequently, one obtains the effective Hamiltonian expressed in the Luttinger liquid formalism. Taking into account Eq. (\ref{eq: scaling}) to keep the leading order, $\hat{H}_n$ can be written as:
\begin{equation}
\label{eq: HnLL}
    \hat{H}_n \simeq \mathcal{E}_n(\omega_c,g,E_0,0) + \lambda_1^{(n)}\hat{H}_{LL}(\bar{v}^{(n)}_S,\bar{v}^{(n)}_N,\bar{v}^{(n)}_J),
\end{equation}
\begin{figure}[b!]
\includegraphics[width = 1.0\hsize]{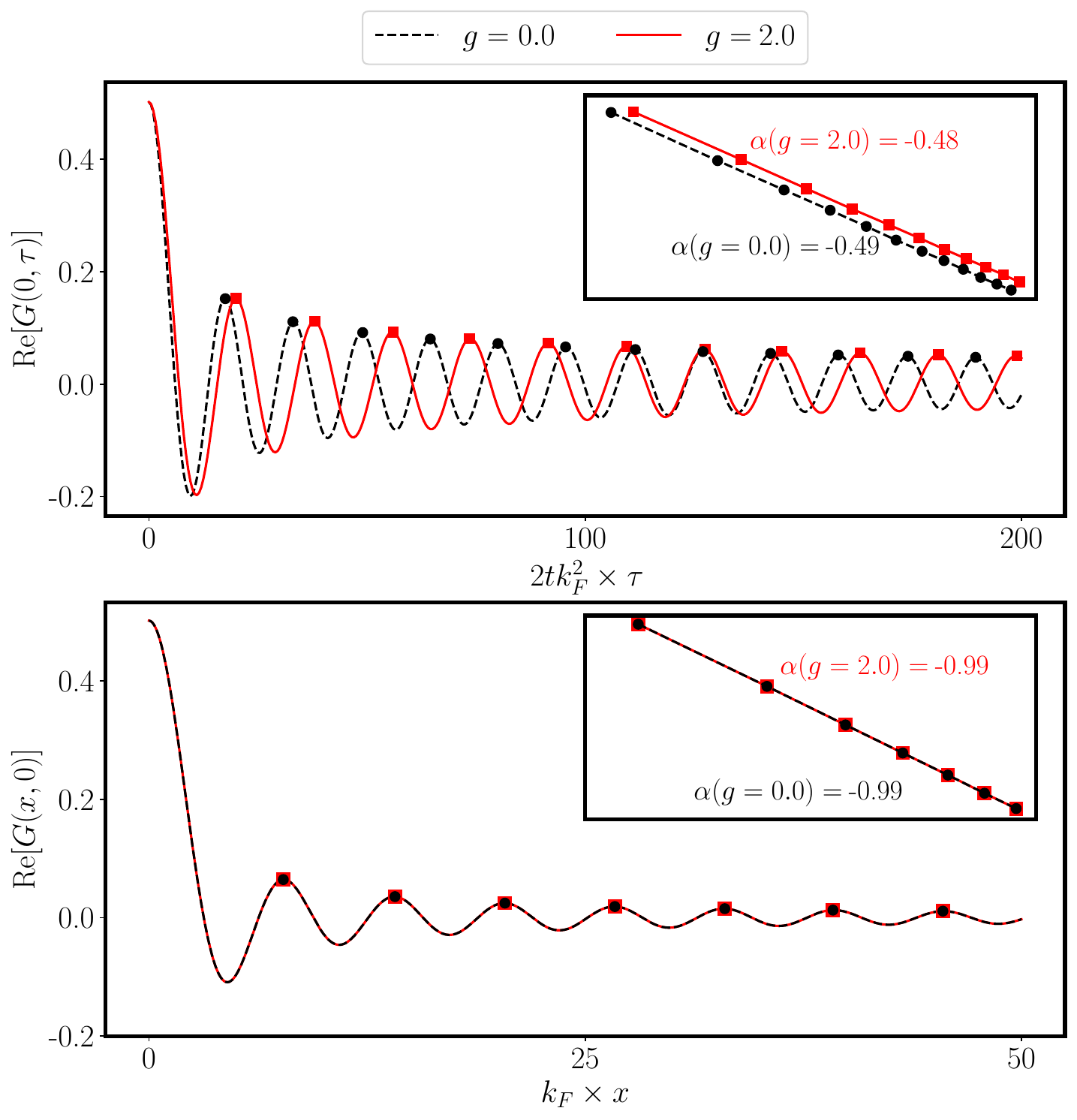}
  \caption{Top panel: Real part of the correlation function $G(0,\tau)$. Bottom panel: Real part of $G(x,0)$. $G(x,\tau)$ measures the propagation of a fermionic excitation between the spacetime points $(0,0)$ and $(x,\tau)$. Both functions are calculated without ($g =0$) and with ($g = 2.0$) light-matter interaction. Inset in both panels display the maxima of the correlation functions on a logarithmic scale, with slope $\alpha(g)$. Other parameters: $N = 500$, $N_e/N = 1/2$, $\omega_c/t = 10$.}
\label{manu_fig: corr}
\end{figure}
where we have factored out the coefficient $\lambda_1^{(n)}$, as the band renormalization effect modifies the entire excitation spectrum. The renormalized velocities induced by the cavity are defined excluding the squeezing effect and are given by:
\begin{equation}
\label{eq: 3 velocities g}
    \begin{aligned}
        \bar{v}^{(n)}_S &= v_F,\\
        \bar{v}^{(n)}_N &= \left(1 + \frac{N\lambda_2^{(n)}}{\lambda_1^{(n)}}\frac{\epsilon_F^2}{\pi v_F}\right)v_F,\\
        \bar{v}^{(n)}_J &= \left(1 + \frac{N\eta_2^{(n)}}{\lambda_1^{(n)}}\frac{v_F}{\pi}\right)v_F.
    \end{aligned}
\end{equation}
In Fig. \ref{manu_fig: 3v}, we compare the results from Eq. (\ref{eq: 3 velocities g}) with the exact tight-binding calculation given by Eq. (\ref{eq: Hk}) for large system. First, the effective model accurately captures the cavity-induced effect. Second, the curve for $\bar{v}_N^{(n)}$ lies slightly below that of $\bar{v}_S^{(n)}$ due to the contribution from $\lambda_2^{(n)}$; this discrepancy disappears when the system is half-filled, i.e., $\epsilon_F = 0$. Third, $\bar{v}_J^{(n)}$ decreases rapidly as a function of $g$ compared to $\bar{v}_N^{(n)}$, reflecting the dominance of $N\eta_2^{(n)}/\lambda_1^{(n)}$ over $N\lambda_2^{(n)}/\lambda_1^{(n)}$ as indicated in Eq. (\ref{eq: scaling}). Expanding the cosine and sine functions in Hamiltonian (\ref{eq: Hp}) to calculate the coefficients analytically and rescaling $g \rightarrow g/\sqrt{N}$, $\bar{v}_J^{(n)}$ recovers the same expression as the Drude weight reported in Ref. \cite{Eckhardt2022}. The suppression of the Drude weight has been experimentally observed \cite{ParaviciniBagliani2018}, with several theoretical efforts to explain it
\cite{Rokaj2022,Eckhardt2022}. Within this framework described by Eq. (\ref{eq: Hp}), for $E<0$ we define $E' = \sqrt{E^2 + I^2}$ and $\text{tan }\varphi = -I/E$ so that the interacting term can be written as $-E'\text{cos} [g (\hat{a}+\hat{a}^{\dagger}) + \varphi]$. Then, Eq. (\ref{eq: 3 velocities g}) directly gives:
\begin{equation}
    \lambda_1^{(n)}\bar{v}_J^{(n)} = \frac{\pi}{N}\frac{\partial^2 \mathcal{E}_n}{\partial \varphi^2}\left(\omega_c,g,E' = \vert E_0\vert,\varphi=0\right).
\end{equation}
In the strong coupling limit, $\varphi$ is dominated by the matrix elements of $g(\hat{a} + \hat{a}^{\dagger})$, making $\mathcal{E}_n$ only weakly sensitive to its fluctuations and consequently suppressing $\bar{v}^{(n)}_J$. 

As a result of this rapid suppression, the usual relation Eq. (\ref{eq: equality})  no longer holds, unlike conventional interacting liquids \cite{Haldane1980}. Moreover, $K = \sqrt{v_J/v_N}$ in conventional Luttinger liquids characterizes the decay of the single-particle correlation function $G(x,\tau) = \langle \hat{c}^{\dagger}(x,\tau) \hat{c}(0,0)\rangle$ \cite{Giamarchi2003,Sutherland2004}. However, an exact evaluation of $G(x,\tau)$ from Eq. (\ref{eq: micro H}) reveals no significant modification of this scaling behavior in the presence of light-matter coupling. In particular, as shown in the top panel of Fig. \ref{manu_fig: corr}, despite light-matter interactions, $G(0,\tau) \sim \vert \tau \vert^{-1/2} $, retaining the same scaling exponent $\alpha = -1/2$ expected from non-interacting, nonlinear Luttinger liquids \cite{Pereira2012}. The visible difference in the oscillation frequency originates from renormalized sound velocity $\bar{v}_S^{(n)}$. Similarly, the spatial correlation $G(x,0) \sim \vert x\vert^{-1}$ presented in the bottom panel remains unchanged with and without the cavity, exhibiting scaling factor $\alpha = -1$ as characteristic of non-interacting fermions. This invariance arises because the calculation depends solely on the ground-state properties, which remain a non-interacting Fermi sea even under light–matter coupling. Therefore, although Eq. (\ref{eq: HnLL}) bears a formal resemblance to an Luttinger liquids Hamiltonian, the cavity-embedded one-dimensional chain must be regarded as a strange Luttinger liquid. 

To conclude this Section, we briefly discuss the eigenstates of Eq. (\ref{eq: HnLL}). The light-matter eigenstates take the following form:
\begin{equation}
    \label{eq: eigen}
    \vert \Psi \rangle = \vert \{m_q\},\Delta N_e,J\rangle \vert \phi^0_n (\delta E, \delta I)\rangle,
\end{equation}
where $\vert m_q\rangle$ denotes the occupation number of the bosonic state $q$ corresponding to $\hat{b}^{\dagger}_q \hat{b}_q$. The states $\vert \phi^0_n (\delta E,\delta I)\rangle$ are the eigenstates of Eq. (\ref{eq: Hp}) at $E = E_0 +\delta E$, $I = \delta I$, in which $\delta E = \sum_q m_q v_F \vert q \vert + \epsilon_F \Delta N_e + \pi v_F/(2N)(\Delta N_e^2 + J^2)$ and $\delta I = -v_F J - \sum_q m_q \epsilon_F \vert q \vert$, correspond to the additional kinetic and current energies due to quantum numbers $\{m_q\}, \Delta N_e, J$. This expression will play an important role in Sec. \ref{sec: ED}, where we evaluate the effects of electron-electron interactions.

\section{Interacting System: Exact Diagonalization and Strange Luttinger Liquid Physics}
\label{sec: ED} 
This Section will consider how the physics of the strange Luttinger liquid is affected by electron-electron interactions. In this regime, the occupation numbers $\{n_k\}$ are no longer conserved quantities, and the corresponding eigenstates cannot generally be written in a factorized form. Furthermore, owing to the photon-mediated long-range interaction, the bosonization approach employed in Ref. \cite{Vadimov2021} is no longer directly applicable. In the present work, however, we focus exclusively on results obtained via exact exact diagonalization (ED), following approaches previously adopted for related light-matter systems in Refs. \cite{Nguyen2024,Becerra2025}. Since Eq. (\ref{eq: micro H}) conserves the number of particles, the Hilbert space can be reduced by restricting to states with $N_e$ particles. The corresponding light-matter energies and eigenstates can then be computed exactly. Then we discuss the fate of the developed strange Luttinger liquids framework in the presence of electron-electron interactions. 

\subsection{ED results: Phase diagram}
\begin{figure}[b!]
\includegraphics[width = 1.0\hsize]{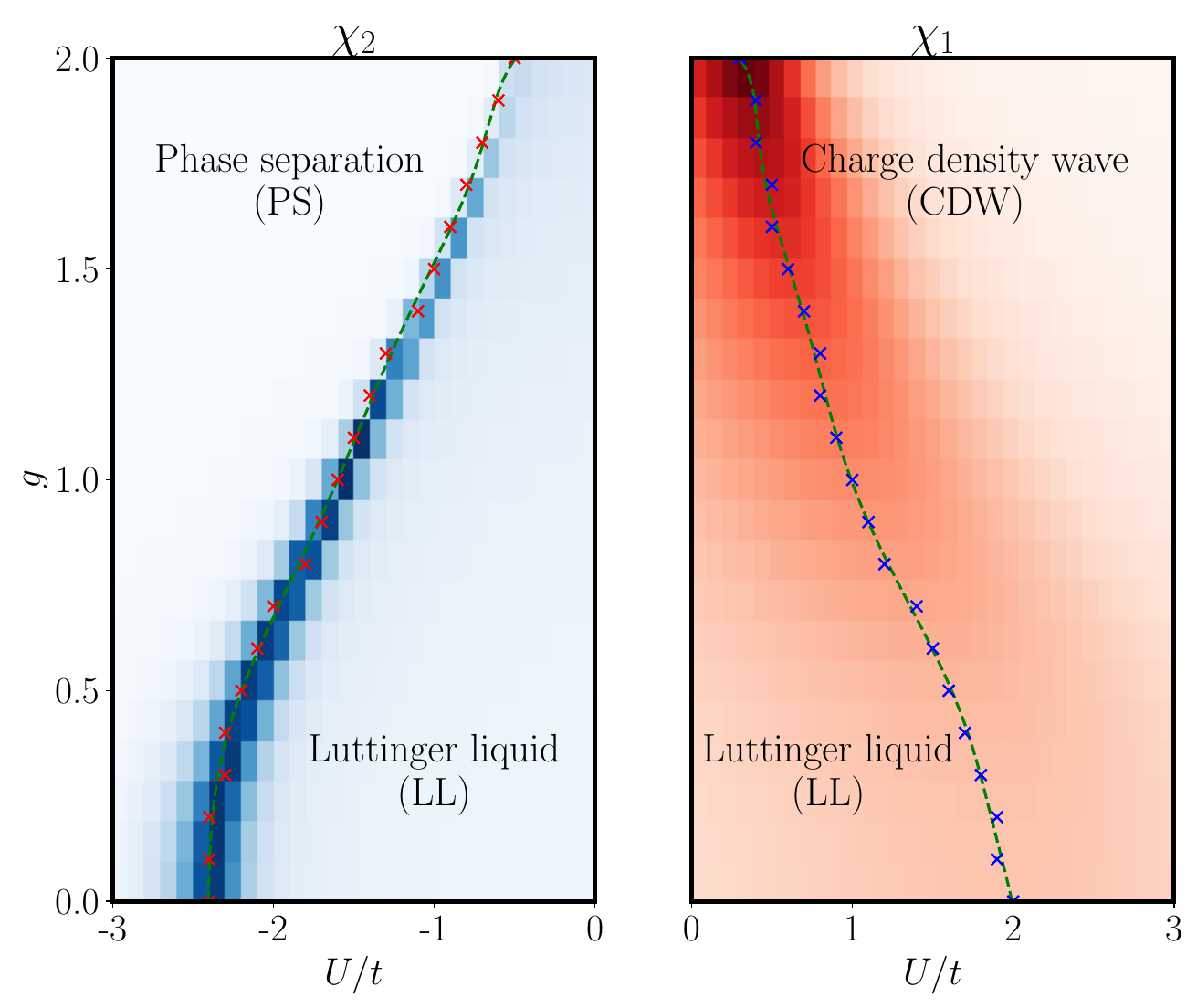}
  \caption{Electronic phase diagram at half-filling, plotted as a function of $U/t$ and $g$ for $N = 12$, $\omega_c/t = 10$. Left panel shows $\chi_2$ for LL-PS transition, while right panel shows $\chi_1$ for LL-CDW transition. In both panels, the cross markers indicate the numerical maxima, while the green dashed lines represent extrapolated results.} 
\label{manu_fig: Phase}
\end{figure}
A spinless chain at half-filling exhibits multiple phases depending on the ratio $U/t$ \cite{Haldane1980,Giamarchi2003,Sutherland2004,Vadimov2021} (see Fig. \ref{manu_fig: Phase}). For $\vert U \vert < 2t$, the system is effectively described by a gapless Luttinger liquid (LL) with a linear dispersing excitation spectrum. When $U > 2t$, the repulsive interaction between two neighboring sites dominates over the kinetic term, leading to an alternating occupation pattern in real space, for example $\vert 1010...\rangle$. This corresponds to a charge density wave (CDW) phase, or equivalently an antiferromagnetic phase under a mapping to the spin-$\frac{1}{2}$ anisotropic XXZ model, with a two-fold degenerate ground state. In contrast, for attractive interactions $U < -2t$, electrons cluster together, resulting in phase separation (PS), where regions of occupied and empty sites are formed, $\vert 11..10..00\rangle$, and the ground state is $N$-fold degenerate in case of periodic boundary conditions, or $2$-fold for open boundary. It should be noted that, since the particle number is fixed, this case does not correspond to a ferromagnetic phase, in which all sites would be either fully occupied or completely empty.

To probe these phase transitions, the system $\hat{H}$ is perturbed by an external source $-\mu\hat{X}$, where $\mu$ is the applied field and $\hat{X}$ depends on the transition of interest. For the LL-CDW transition, we choose $\hat{X} = \hat{X}_1 = \sum_{n}(-1)^n\hat{c}^{\dagger}_n\hat{c}_n$, which breaks the degeneracy by favoring the ground state in which even sites are occupied. For the LL-PS transition, $\hat{X} = \hat{X}_2 = \hat{c}^{\dagger}_1\hat{c}_1$, selecting clusters in which the first site is occupied. The corresponding susceptibilities are then defined as $\chi_{i} = \lim_{\mu\to 0^+}\partial_{\mu}\langle\hat{X}_i\rangle$, where $\langle \hat{X}_i\rangle$ is calculated from the exact ground states of $\hat{H} - \mu \hat{X}_i$.

Without light-matter coupling, we report that $\chi_1$ ($\chi_2$) has a peak around $U = +2t$ ($U = -2t$), which is coherent with the predicted results \cite{Giamarchi2003,Sutherland2004}. The phase boundaries are determined from the maxima of $\chi_i$. Fig. \ref{manu_fig: Phase} shows the phase diagram calculated for $N=12$ at half-filling, where in both transitions, weaker $U$ is required to induce phase transition in the presence of light-matter coupling. This behavior can be understood from the cavity-induced renormalization of the hopping amplitude $t$, which effectively increases the ratio $U/t$ governing the phase transitions.\\
\begin{figure}[t!]
\includegraphics[width = 1.0\hsize]{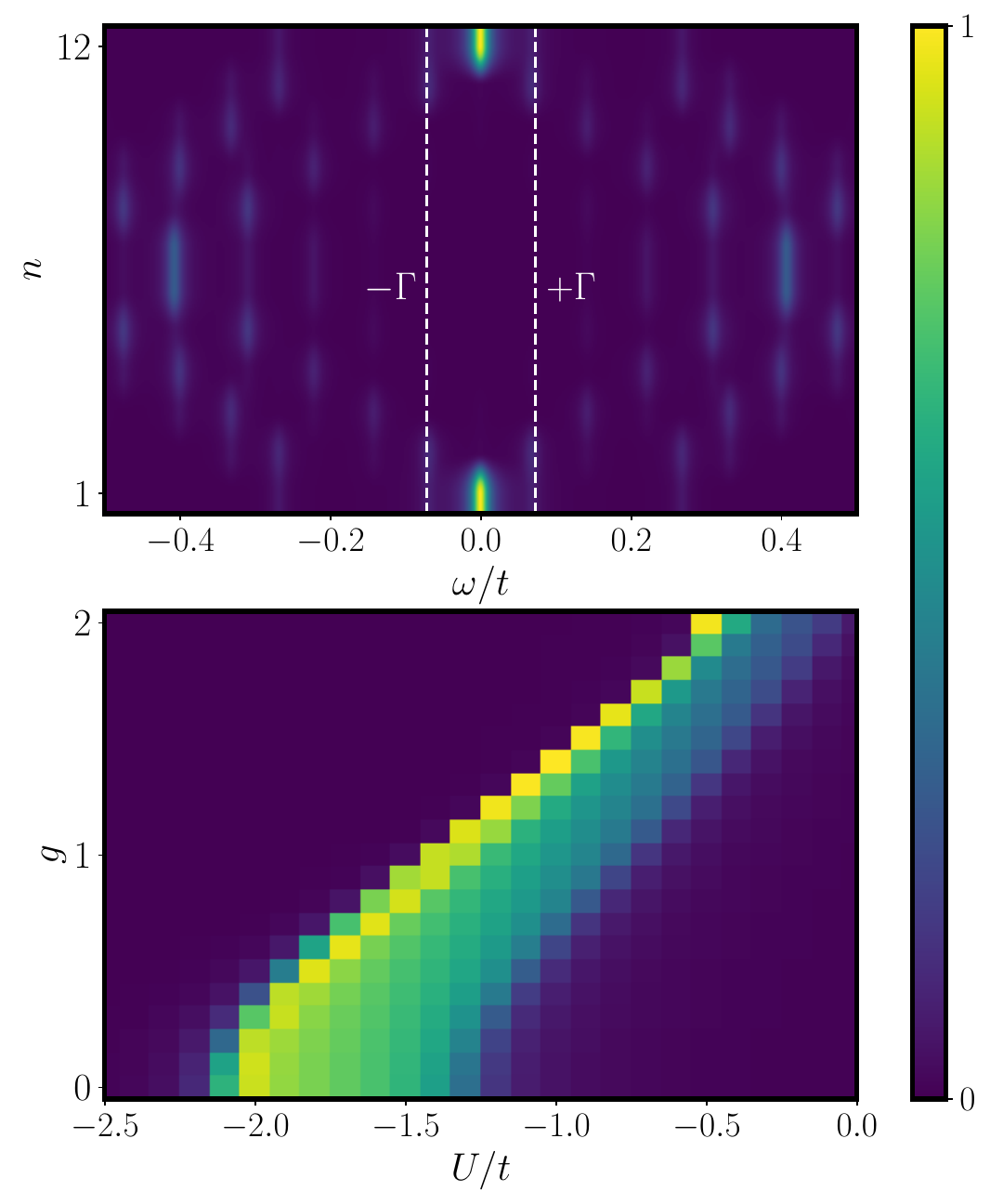}
  \caption{Top: Local density of states $A_n(\omega)$ at half-filling for $U/t = -2$ in the absence of light-matter coupling. Bottom: Edge spectral weight $W(U,g)$ at half-filling as a function of $U/t$ and $g$. Other parameters: $N = 12$, $\omega_c/t = 10$, $\Gamma = 0.3 \times$ level spacing $\pi t/N$. The spectral functions are shown in arbitrary units, and are normalized to their respective maxima.}
\label{manu_fig: weight}
\end{figure}
\subsection{ED results: Majorana-like zero modes}
In the regime $-2t < U < 0$, Ref. \cite{Vadimov2021} reports the emergence of localized zero modes near the edges when open boundary conditions are applied to Eq. (\ref{eq: micro H}) for $g = 0$ (see the top panel of Fig. \ref{manu_fig: weight}). In the presence of light-matter interaction, to probe the presence of Majorana-like zero modes, we measure the local density of states $A_n(\omega)$ at site $n$, defined by:
\begin{equation}
    \begin{aligned}
    A_n(\omega) &= \sum_{\mu} \vert \langle \Psi_{\mu}\vert\hat{c}^{\dagger}_n\vert\Psi_0\rangle\vert^2\delta(\omega - \mathcal{E}_{\mu} + \mathcal{E}_0)\\
    &+ \sum_{\nu} \vert \langle \Psi_{\nu}\vert\hat{c}_n\vert\Psi_0\rangle\vert^2\delta(\omega - \mathcal{E}_{0} + \mathcal{E}_{\nu}).
    \end{aligned}
\end{equation}
In the above expression, $\vert \Psi_{\mu}\rangle$ ($\vert \Psi_{\nu}\rangle$) are eigenstates of $\hat{H}$ in the subspace containing $N_e + 1$ ($N_e -1$) particles, with corresponding energies $\mathcal{E}_{\mu}$ ($\mathcal{E}_{\nu}$), while $\vert \Psi_0\rangle$ denotes the ground state with energy $\mathcal{E}_0$. In Fig. \ref{manu_fig: weight}, we quantify the edge spectral weight $W(U,g) = \int_{-\Gamma}^{+\Gamma} d\omega (A_1(\omega) + A_N(\omega))$ within an energy window $\Gamma$ smaller than the finite-size level spacing $\pi/N$. We observe that $W$ is maximized near the phase-transition points, and the finite light-matter coupling $g$ induces shifts similar to those seen in the LL-PS phase transition.

\subsection{Strange Luttinger liquid description}
We now incorporate two-particle interactions into Hamiltonian (\ref{eq: HnLL}) and argue the extension of the strange Luttinger liquid framework. Recalling that $\hat{\rho}_{q,\sigma} = \sum_{\vert k - \sigma k_F\vert < \Lambda}\hat{c}^{\dagger}_{k+q}\hat{c}_{k}$, the interaction term can be expressed in LL formalism as \cite{Haldane1980,Giamarchi2003,Sutherland2004}:
\begin{equation}
    \begin{aligned}
        \hat{V} &= \frac{\pi}{N}\sum_{\sigma,q}\left(V_{1,q}\hat{\rho}_{q,\sigma}\hat{\rho}_{-q,\sigma} + V_{2,q}\hat{\rho}_{q,\sigma}\hat{\rho}_{-q,-\sigma}\right),
    \end{aligned}
\end{equation}
where $V_{1,q}$ ($V_{2,q}$) denotes the interaction strength for generating two particle-hole excitations around one (two) Dirac points. First, $\hat{V}$ conserves particle number, so $\Delta N_e$ remains a good quantum number. Second, the limit $q \sim 0$ implies that particle-hole excitations are confined to the same Dirac point. Under this approximation, the particle number at each Dirac point is conserved, and $J$ also remains a good quantum number. We now evaluate the matrix elements of $\hat{V}$ between the light-matter eigenstates $\vert \Psi'\rangle$  and $\vert \Psi\rangle$ given in Eq. (\ref{eq: eigen}). Without any approximation, $\langle \Psi'\vert\hat{V}\vert\Psi\rangle $ takes the form:
\begin{equation}
\label{eq: matrix elements}
    \begin{aligned}
        \langle \Psi'\vert\hat{V}\vert\Psi\rangle  = &\langle \{m'_q\},\Delta N'_e,J'\vert \hat{V}\vert \{m_q\},\Delta N_e,J\rangle\\
        &\times \langle \phi^0_{n'} (\delta E', \delta I')\vert \phi^0_n (\delta E, \delta I)\rangle.
    \end{aligned}   
\end{equation}
The interaction term couples different $\hat{H}_n$, with the corresponding form factor given by the second line of Eq. (\ref{eq: matrix elements}). In general, this form factor depends nonlinearly on both $\delta E$ and $\delta I$, thereby generating light-matter entanglement and preventing a simple effective Hamiltonian. Nevertheless, useful approximations can be made. First, in the large $N$ limit, the second line of Eq. (\ref{eq: matrix elements}) reduces to $\delta_{n'n} + \mathcal{O}(N^{-3/4})$ (see the \SM). Consequently, matrix elements connecting different photonic sectors vanish asymptotically with increasing system size. Second, since $\hat{V}$ couples electronic states that differ by at most two particle-hole excitations, the corresponding coupled light-matter states remain energetically well-separated when $8t < \omega_c$. We have numerically verified that the correction of the form factor has the order of $10^{-4}$ for a system with $N = 500$ and $\omega_c/t = 10$. Combining these two approximations, photonic sectors effectively decouple, allowing $\hat{V}$ to be incorporated perturbatively into $\hat{H}_n$.

Expressing $\hat{\rho}_{q,\sigma}$ as a linear combination of $\hat{b}_q^{\dagger}$ and $\hat{b}_q$ \cite{Haldane1980}, the combined Hamiltonian $\hat{H}_n + \hat{V}$ becomes quadratic in bosonic operators. After performing a Bogoliubov transformation, the resulting Hamiltonian $\tilde{H}_n = \hat{H}_n + \hat{V}$ can be written as:
\begin{equation}
\label{eq: Heff U g}
    \begin{aligned}
    \tilde{H}_n &\simeq \mathcal{E}_n(\omega_c,g,E_0,0) + \frac{\lambda_1^{(n)}}{2}\sum_q\vert q \vert(\tilde{v}_S^{(n)}-v_F)\\
    &+ \lambda_1^{(n)}\tilde{H}_{LL}(\tilde{v}^{(n)}_S,\tilde{v}^{(n)}_N,\tilde{v}^{(n)}_J),
    \end{aligned}
\end{equation}
where $\tilde{H}_{LL}$ has the same structure as $\hat{H}_{LL}$ in Eq. (\ref{eq: TLL}) but is expressed in terms of the dressed bosonic operators $\hat{\beta}^{\dagger}_q$ and $\hat{\beta}_q$. Using $V_{i,0} = U/\pi$ for both $i=1$ and $2$, the dressed velocities are:\\
\begin{equation}
    \label{eq: rev}
    \begin{aligned}
        &\tilde{v}^{(n)}_S =  \sqrt{ v_F^2 + v_F\frac{2U}{\pi\lambda_1^{(n)}}},\\
        &\tilde{v}^{(n)}_N = v_F + \epsilon_F^2\frac{N\lambda_2^{(n)}}{ \pi\lambda_1^{(n)}} + \frac{2U}{\pi\lambda_1^{(n)}},\\
        &\tilde{v}^{(n)}_J = v_F + v_F^2\frac{N\eta_2^{(n)}}{ \pi\lambda_1^{(n)}}.
    \end{aligned}
\end{equation}
Due to second-order corrections, $\tilde{v}_J^{(n)}$ is always strongly suppressed, and we generally have $\tilde{v}_S^{(n)} \neq \sqrt{\tilde{v}_J^{(n)}\tilde{v}_N^{(n)}}$. Moreover, since $\lambda_1^{(n)} < 1$, electron-electron interactions are effectively enhanced in the presence of light-matter coupling, which accounts for the phase diagram obtained via exact diagonalization in Fig. \ref{manu_fig: Phase}. Eq. (\ref{eq: rev}) also demonstrates that cavity-induced phase transitions require nonzero $U$. The results are here are perturbative in $U$. The exact diagonalization calculations have not the resolution to probe these effects due to the finite-size effects.

\section{Conclusion \& Outlook}
\label{sec: conclusion}
We have investigated a one-dimensional interacting electronic chain coupled to a homogeneous quantized cavity field and demonstrated that cavity vacuum fluctuations qualitatively modify its low-energy physics. In the absence of electron--electron interactions, we derived an effective low-energy theory in which each photonic sector is described by a Luttinger-like Hamiltonian with cavity-renormalized parameters. While this theory retains the formal structure of conventional Luttinger liquid theory, it violates the exact relation between the sound, density, and current velocities that underlies Luttinger universality. We therefore identify this regime as a \emph{strange Luttinger liquid}, where the cavity field selectively renormalizes current and density responses without modifying the ground-state correlation exponents.

For finite nearest-neighbor interactions, exact diagonalization reveals that the cavity field substantially reshapes the phase diagram by renormalizing the effective kinetic energy. In particular, the critical interaction strengths separating the Luttinger liquid, charge-density-wave, and phase-separated phases are significantly shifted by the light--matter coupling. We also showed that the spectral signatures of Majorana-like edge modes are modified by the cavity field, demonstrating that vacuum fluctuations provide a new mechanism for controlling topological many-body features in number-conserving interacting systems.

Finally, we proposed an extension of the strange Luttinger liquid framework to the interacting regime. Although our analytical treatment is perturbative in the interaction strength, it provides a unified low-energy description that consistently accounts for both cavity-induced renormalization and electron--electron correlations, in agreement with the trends observed in exact diagonalization. Our results establish cavity quantum electrodynamics as a route toward engineering unconventional one-dimensional quantum liquids whose low-energy properties lie beyond the standard Luttinger liquid paradigm.

\acknowledgements
We acknowledge financial support from the French agency ANR through the project CaVdW (ANR-21-CE30-0056-0) and a grant (Polaritonic) from the French Government managed by the
ANR under the France 2030 programme with the reference ANR-24-RRII-0001. 

\appendix
\makeatletter
\renewcommand{\@seccntformat}[1]{\thesection\hspace{1em}}  
\makeatother

\section{Derivation of the Effective Coupling Coefficients}
We study a cavity-embedded tight-binding chain with $N$ sites and periodic boundary conditions. After projecting the full Hamiltonian onto a generic occupied electronic configuration $\vert \psi\rangle = \vert  \{ n_k \} \rangle$ characterized by its kinetic and current energies $E$, $I$, photon sector is governed by an effective Hamiltonian $\hat{H}_0^{(p)}(\omega_c,g,E,I)$:
\begin{equation}
\label{SM: eq-H}
    \hat{H}_0^{(p)} = \omega_c \hat{a}^{\dagger}\hat{a} + \text{cos}\left[g(\hat{a}+\hat{a}^{\dagger})\right]E + \text{sin}\left[g(\hat{a} +\hat{a}^{\dagger})\right]I.
\end{equation}
With eigenstates $\mathcal{E}_n(\omega_c, g, E, I)$. Now we restrict our attention to small excitations above the non-interacting ground state $\vert \psi_0 \rangle$, whose kinetic and current energies $E_0$ and $I_0 = 0$. Expanding $\mathcal{E}_n(\omega_c, g, E, I)$ about $E_0$ and $I_0$ to second order yields $\mathcal{E}_n(\omega_c,g,E,I) \simeq \mathcal{E}_n(\omega_c,g,E_0,0) + \lambda_1^{(n)}(E -E_0) + \lambda_2^{(n)} (E-E_0)^2/2 + \eta_2^{(n)}I^2/2$, where the expansion coefficients are defined by:
\begin{equation}
\label{eq-sm: 4 coeffs}
    \begin{aligned}
        &\lambda_1^{(n)} = \frac{\partial \mathcal{E}_n}{\partial E}(\omega_c,g,E_0,0), \: \lambda_2^{(n)}= \frac{\partial^2\mathcal{E}_n}{\partial E^2}(\omega_c,g,E_0,0),\\
        &\eta_1^{(n)} = \frac{\partial \mathcal{E}_n}{\partial I}(\omega_c,g,E_0,0), \: \eta_2^{(n)}= \frac{\partial^2\mathcal{E}_n}{\partial I^2}(\omega_c,g,E_0,0).
    \end{aligned}
\end{equation}
The derivatives can be evaluated exactly by applying the Feynman-Hellmann theorem. Let $\{\vert \phi^0_{m}\rangle\}$ denote the eigenbasis of Hamiltonian (\ref{SM: eq-H}) calculated at the reference point $(\omega_c,g,E_0,0)$, corresponding to energies $\{\mathcal{E}_n^0\}$. We introduce the matrix elements $C_{mn}(g) = \langle \phi^0_{m}\vert \text{cos}\left[g(\hat{a} +\hat{a}^{\dagger})\right]\vert \phi^0_{n}\rangle$, $S_{mn}(g) = \langle \phi^0_{m}\vert \text{sin}\left[g(\hat{a} +\hat{a}^{\dagger})\right]\vert \phi^0_{n}\rangle$, from which one obtains:
\begin{equation}
    \begin{aligned}
        &\lambda_1^{(n)} = C_{nn}(g),\: \lambda_2^{(n)} = 2\sum_{m \neq n}\frac{\vert C_{mn}(g)\vert^2}{\mathcal{E}^0_{n}-\mathcal{E}^0_{m}},
    \end{aligned}
\end{equation}
while the coefficients $\eta_i^{(n)}$ follow from analogous expressions with the substitution $C \rightarrow S$.
To obtain analytical expressions for these coefficients, we expand the cosine and sine functions appearing in Hamiltonian (\ref{SM: eq-H}) to leading order in $g$. Under this approximation, $\hat{H}_0^{(p)}$ becomes quadratic in $\hat{a}$ and $\hat{a}^\dagger$ and can be diagonalized by a Bogoliubov transformation, yielding the dressed operators $\hat{\alpha}$ and $\hat{\alpha}^\dagger$. In this representation one finds:
\begin{equation}
\label{SM: eq-Heff}
\hat{H}_p(\omega_c,g,E,I) \simeq s\omega_c \hat{\alpha}^{\dagger}\hat{\alpha} + E - \frac{g^2I^2}{s^2\omega_c} + \frac{\omega_c(s-1)}{2},
\end{equation}
with $s = \sqrt{1-2g^2E/\omega_c}$, and the dressed operators take the form:
\begin{equation}
    \label{SM: eq-dphot}
    \begin{aligned}
    &\hat{\alpha}^{\dagger} = \frac{1}{2\sqrt{s}}\left[(s-1)\hat{a} + (s+1)\hat{a}^{\dagger} + \frac{2gI}{s\omega_c}\right],\\ &\hat{\alpha}^{\dagger} + \hat{\alpha} = \sqrt{s}(\hat{a}^{\dagger}+\hat{a}) + \frac{2gI}{s^{3/2}\omega_c}.     
    \end{aligned}
\end{equation}
Using Eqs. (\ref{SM: eq-Heff}) and (\ref{SM: eq-dphot}), the states $\vert \phi^0_m\rangle$ becomes the dressed number states $\vert \tilde{m}\rangle$ corresponding to $\hat{\alpha}$. For $E = E_0 = -Nv_F/\pi$ and $I = I_0 = 0$, the coefficients are:
\begin{equation}
\label{SM: eq-acoeff}
    \begin{aligned}
        \lambda_1^{(n)} &\simeq 1 - \frac{\left(n + \frac{1}{2}\right)g^2}{\sqrt{1 + \frac{2g^2 N v_F}{\pi\omega_c}}},\\ \lambda_2^{(n)} &\simeq \frac{-\left(n + \frac{1}{2}\right)g^4}{\omega_c\left(1 + \frac{2g^2 N v_F}{\pi\omega_c} \right)^{3/2}},\\
        \eta_2^{(n)} &\simeq \frac{-2g^2}{\omega_c \left(1 + \frac{2g^2 N v_F}{\pi\omega_c}\right)}.
    \end{aligned}
\end{equation}
and $\eta_1^{(n)} = 0$. From Eq. (\ref{SM: eq-acoeff}), one finds the scalings $\lambda_2^{(n)}/\lambda_1^{(n)} = \mathcal{O}(N^{-3/2})$, $\eta_2^{(n)}/\lambda_1^{(n)}  = \mathcal{O}(N^{-1})$.

\section{Bosonization of the Kinetic and Current Operators}
After fixing a reference state $\vert \psi_0\rangle$ containing $N_0$ electrons, the Hamiltonian of the free chain can be bosonized and diagonalized in the basis $\vert \{n_q\}, N_e, J\rangle$, where $n_q$ denotes number states of  bosonic excitations $\hat{b}^{\dagger}_q$, $\Delta N_e = N_e-N_0$ counts the additional electrons, and $J$ measures the imbalance in terms of particles between $+k_F$ and $-k_F$, corresponding to the persistent current. In terms of the quantum numbers $N_e$ and $J$, the kinetic energy reads:
\begin{equation}
    \hat{T} \simeq E_0 + v_F\sum_q \vert q \vert \hat{b}^{\dagger}_q\hat{b}_q + \epsilon_F \Delta N_e + \frac{\pi v_F}{2N} \Delta N_e^2 + \frac{\pi v_F}{2N} J^2.
\end{equation}
To express the current operator $\hat{I}$ in Luttinger formalism, we introduce a flux $\phi$ through the chain. In the presence of a vector potential $A = \phi/N$, $\hat{T}$ is written in term of fermionic fields as:
\begin{equation}
    \hat{T} \simeq \frac{v_F}{2\pi}\int dx \left\{\left[\pi \Pi(x) - \frac{\phi}{N}\right]^2 + \left[\nabla\Phi(x)\right]^2\right\},
\end{equation}
with the field $\Phi(x)$ and its canonical conjugate momenta $\Pi(x)$. The current operator is then given by:
\begin{equation}
    \hat{I} = N\frac{\partial \hat{T}}{\partial \phi}\Big{|}_{\phi=0} \simeq -v_F\int dx\: \Pi(x). 
\end{equation}
The momentum field can be expressed in terms of bosonic operators as:
\begin{equation}
    \Pi (x) = \frac{J}{N} + \frac{1}{N}\sum_{q \neq 0}\sqrt{\frac{N \vert q \vert}{2\pi}}\text{sign}(q) e^{-\alpha\vert q\vert/2 - iqx}\left(\hat{b}^{\dagger}_q - \hat{b}_{-q}\right),
\end{equation}
with $\alpha \rightarrow 0$. After integrating over $x$, the oscillatory vanish, leaving only the zero-mode contribution. Hence, the current operator reduces to:
\begin{equation}
    \hat{I} = N\frac{\partial \hat{T}}{\partial \phi}\Big{|}_{\phi=0} \simeq -v_F J.
\end{equation}

\begin{widetext}
\section{Overlap between photon eigenstates}
Let $\vert \phi^0_{n} (\delta E, \delta I)\rangle$ denote an eigenstate of Hamiltonian (\ref{SM: eq-H}) evaluated at the shifted parameters $(\omega_c,g,E_0 + \delta E, \delta I)$. We seek to express it as a linear combination of the unperturbed eigenstates $\vert \phi^0_{m}(0,0)\rangle = \vert \phi^0_{m}\rangle$ of the ground state, corresponding to energies $\mathcal{E}^0_m$. To first order, perturbation theory yields:
\begin{equation}
    \vert \phi^0_{n}(\delta E, \delta I)\rangle \simeq \vert \phi^0_{n}\rangle +  \sum_{m\neq n}\frac{C_{mn}(g)\delta E + S_{mn}(g)\delta I}{\mathcal{E}^0_{n}-\mathcal{E}^0_{m}}\vert\phi^0_{m}\rangle.
\end{equation}
It in turn gives the overlap:
\begin{equation}
\label{eq-overlap_exact}
    \begin{aligned}
    \langle \phi^0_{n'}(\delta E',\delta I')\vert \phi^0_{n}(\delta E,\delta I)\rangle &\simeq \delta_{n'n} + (1 - \delta_{n'n})\frac{C_{n'n}(g)(\delta E' - \delta E) + S_{n'n}(g)(\delta I' - \delta I)}{\mathcal{E}^0_{n'}-\mathcal{E}^0_{n}}\\
    &- \sum_{m \neq n',n}\frac{\left[C_{n'm}(g)\delta E' + S_{n'm}(g)\delta I'\right]\left[C_{mn}(g)\delta E + S_{mn}(g)\delta I \right]}{(\mathcal{E}^0_{n'}-\mathcal{E}^0_{m})(\mathcal{E}^0_{m}-\mathcal{E}^0_{n})}.
    \end{aligned}
\end{equation}
To proceed analytically, we expand the sine and cosine functions in Eq. (\ref{SM: eq-H}) to leading order. Introducing $s_0 = \sqrt{1 - 2g^2E_0/\omega_c} = \sqrt{1 + 2g^2Nv_F/(\pi \omega_c)}$, we obtain:
\begin{equation}
\label{SM: entanglement}
    \begin{aligned}
        \langle \phi^0_{n'}(\delta E',\delta I') \vert \phi^0_{n}(\delta E,\delta I)\rangle &\simeq \delta_{n'n} + (\delta I' - \delta I)\frac{g}{s_0^{3/2}\omega_c}\left[\sqrt{n+1}\delta_{n',n+1} - \sqrt{n}\delta_{n',n-1}\right]\\
        + (\delta E' - \delta E)&\frac{g^2}{4s_0^2\omega_c}\left[\sqrt{(n+2)(n+1)}\delta_{n',n+2} - \sqrt{n(n-1)}\delta_{n',n-2}\right] + \mathcal{O}(s_0^{-3}).
    \end{aligned}
\end{equation}
Since the corrections scale as $\mathcal{O}(N^{-3/4})$, even the first-order contributions can be safely neglected in the large $N$ limit.
\end{widetext}

\bibliography{Bib.bib}
	
\end{document}